\pdfoutput=1
\documentclass[11pt,twoside,a4paper,pdftex,cmspaper,final,collab]{cms-tdr}

\begin{document}\cmsNoteHeader{CFT-09-005}
%
%
%

%
%
\hyphenation{env-iron-men-tal}
\hyphenation{had-ron-i-za-tion}
\hyphenation{cal-or-i-me-ter}
\hyphenation{de-vices}
%
%
\hyphenation{env-iron-men-tal}
\hyphenation{had-ron-i-za-tion}
\hyphenation{cal-or-i-me-ter}
\hyphenation{de-vices}
\global\emergencystretch = .9\hsize
%
%
\RCS$Revision: 1.66 $
\RCS$Date: 2010/02/16 17:31:04 $
\RCS$Name:  $
\providecommand{\MyGEANTfour} {{\textsc{{\Large g}eant4}}\xspace}
%
%
%

\newcommand {\etal}{\mbox{et al.}\xspace} 
\newcommand {\ie}{\mbox{i.e.}\xspace}     
\newcommand {\eg}{\mbox{e.g.}\xspace}     
\newcommand {\etc}{\mbox{etc.}\xspace}     
\newcommand {\vs}{\mbox{\sl vs.}\xspace}      
\newcommand {\mdash}{\ensuremath{\mathrm{-}}} 

\newcommand {\Lone}{Level-1\xspace} 
\newcommand {\Ltwo}{Level-2\xspace}
\newcommand {\Lthree}{Level-3\xspace}

\providecommand{\ACERMC} {\textsc{AcerMC}\xspace}
\providecommand{\ALPGEN} {{\textsc{alpgen}}\xspace}
\providecommand{\CHARYBDIS} {{\textsc{charybdis}}\xspace}
\providecommand{\CMKIN} {\textsc{cmkin}\xspace}
\providecommand{\CMSIM} {{\textsc{cmsim}}\xspace}
\providecommand{\CMSSW} {{\textsc{cmssw}}\xspace}
\providecommand{\COBRA} {{\textsc{cobra}}\xspace}
\providecommand{\COCOA} {{\textsc{cocoa}}\xspace}
\providecommand{\COMPHEP} {\textsc{CompHEP}\xspace}
\providecommand{\EVTGEN} {{\textsc{evtgen}}\xspace}
\providecommand{\FAMOS} {{\textsc{famos}}\xspace}
\providecommand{\GARCON} {\textsc{garcon}\xspace}
\providecommand{\GARFIELD} {{\textsc{garfield}}\xspace}
\providecommand{\GEANE} {{\textsc{geane}}\xspace}
\providecommand{\GEANTfour} {{\textsc{geant4}}\xspace}
\providecommand{\GEANTthree} {{\textsc{geant3}}\xspace}
\providecommand{\GEANT} {{\textsc{geant}}\xspace}
\providecommand{\HDECAY} {\textsc{hdecay}\xspace}
\providecommand{\HERWIG} {{\textsc{herwig}}\xspace}
\providecommand{\HIGLU} {{\textsc{higlu}}\xspace}
\providecommand{\HIJING} {{\textsc{hijing}}\xspace}
\providecommand{\IGUANA} {\textsc{iguana}\xspace}
\providecommand{\ISAJET} {{\textsc{isajet}}\xspace}
\providecommand{\ISAPYTHIA} {{\textsc{isapythia}}\xspace}
\providecommand{\ISASUGRA} {{\textsc{isasugra}}\xspace}
\providecommand{\ISASUSY} {{\textsc{isasusy}}\xspace}
\providecommand{\ISAWIG} {{\textsc{isawig}}\xspace}
\providecommand{\MADGRAPH} {\textsc{MadGraph}\xspace}
\providecommand{\MCATNLO} {\textsc{mc@nlo}\xspace}
\providecommand{\MCFM} {\textsc{mcfm}\xspace}
\providecommand{\MILLEPEDE} {{\textsc{millepede}}\xspace}
\providecommand{\ORCA} {{\textsc{orca}}\xspace}
\providecommand{\OSCAR} {{\textsc{oscar}}\xspace}
\providecommand{\PHOTOS} {\textsc{photos}\xspace}
\providecommand{\PROSPINO} {\textsc{prospino}\xspace}
\providecommand{\PYTHIA} {{\textsc{pythia}}\xspace}
\providecommand{\SHERPA} {{\textsc{sherpa}}\xspace}
\providecommand{\TAUOLA} {\textsc{tauola}\xspace}
\providecommand{\TOPREX} {\textsc{TopReX}\xspace}
\providecommand{\XDAQ} {{\textsc{xdaq}}\xspace}

\newcommand {\DZERO}{D\O\xspace}     


\newcommand{\de}{\ensuremath{^\circ}}
\newcommand{\ten}[1]{\ensuremath{\times \text{10}^\text{#1}}}
\newcommand{\unit}[1]{\ensuremath{\text{\,#1}}\xspace}
\newcommand{\mum}{\ensuremath{\,\mu\text{m}}\xspace}
\newcommand{\micron}{\ensuremath{\,\mu\text{m}}\xspace}
\newcommand{\cm}{\ensuremath{\,\text{cm}}\xspace}
\newcommand{\mm}{\ensuremath{\,\text{mm}}\xspace}
\newcommand{\mus}{\ensuremath{\,\mu\text{s}}\xspace}
\newcommand{\keV}{\ensuremath{\,\text{ke\hspace{-.08em}V}}\xspace}
\newcommand{\MeV}{\ensuremath{\,\text{Me\hspace{-.08em}V}}\xspace}
\newcommand{\GeV}{\ensuremath{\,\text{Ge\hspace{-.08em}V}}\xspace}
\newcommand{\TeV}{\ensuremath{\,\text{Te\hspace{-.08em}V}}\xspace}
\newcommand{\PeV}{\ensuremath{\,\text{Pe\hspace{-.08em}V}}\xspace}
\newcommand{\keVc}{\ensuremath{{\,\text{ke\hspace{-.08em}V\hspace{-0.16em}/\hspace{-0.08em}c}}}\xspace}
\newcommand{\MeVc}{\ensuremath{{\,\text{Me\hspace{-.08em}V\hspace{-0.16em}/\hspace{-0.08em}c}}}\xspace}
\newcommand{\GeVc}{\ensuremath{{\,\text{Ge\hspace{-.08em}V\hspace{-0.16em}/\hspace{-0.08em}c}}}\xspace}
\newcommand{\TeVc}{\ensuremath{{\,\text{Te\hspace{-.08em}V\hspace{-0.16em}/\hspace{-0.08em}c}}}\xspace}
\newcommand{\keVcc}{\ensuremath{{\,\text{ke\hspace{-.08em}V\hspace{-0.16em}/\hspace{-0.08em}c}^\text{2}}}\xspace}
\newcommand{\MeVcc}{\ensuremath{{\,\text{Me\hspace{-.08em}V\hspace{-0.16em}/\hspace{-0.08em}c}^\text{2}}}\xspace}
\newcommand{\GeVcc}{\ensuremath{{\,\text{Ge\hspace{-.08em}V\hspace{-0.16em}/\hspace{-0.08em}c}^\text{2}}}\xspace}
\newcommand{\TeVcc}{\ensuremath{{\,\text{Te\hspace{-.08em}V\hspace{-0.16em}/\hspace{-0.08em}c}^\text{2}}}\xspace}

\newcommand{\pbinv} {\mbox{\ensuremath{\,\text{pb}^\text{$-$1}}}\xspace}
\newcommand{\fbinv} {\mbox{\ensuremath{\,\text{fb}^\text{$-$1}}}\xspace}
\newcommand{\nbinv} {\mbox{\ensuremath{\,\text{nb}^\text{$-$1}}}\xspace}
\newcommand{\percms}{\ensuremath{\,\text{cm}^\text{$-$2}\,\text{s}^\text{$-$1}}\xspace}
\newcommand{\lumi}{\ensuremath{\mathcal{L}}\xspace}
\newcommand{\Lumi}{\ensuremath{\mathcal{L}}\xspace}
%
\newcommand{\LvLow}  {\ensuremath{\mathcal{L}=\text{10}^\text{32}\,\text{cm}^\text{$-$2}\,\text{s}^\text{$-$1}}\xspace}
\newcommand{\LLow}   {\ensuremath{\mathcal{L}=\text{10}^\text{33}\,\text{cm}^\text{$-$2}\,\text{s}^\text{$-$1}}\xspace}
\newcommand{\lowlumi}{\ensuremath{\mathcal{L}=\text{2}\times \text{10}^\text{33}\,\text{cm}^\text{$-$2}\,\text{s}^\text{$-$1}}\xspace}
\newcommand{\LMed}   {\ensuremath{\mathcal{L}=\text{2}\times \text{10}^\text{33}\,\text{cm}^\text{$-$2}\,\text{s}^\text{$-$1}}\xspace}
\newcommand{\LHigh}  {\ensuremath{\mathcal{L}=\text{10}^\text{34}\,\text{cm}^\text{$-$2}\,\text{s}^\text{$-$1}}\xspace}
\newcommand{\hilumi} {\ensuremath{\mathcal{L}=\text{10}^\text{34}\,\text{cm}^\text{$-$2}\,\text{s}^\text{$-$1}}\xspace}


\newcommand{\zp}{\ensuremath{\mathrm{Z}^\prime}\xspace}


\newcommand{\kt}{\ensuremath{k_{\mathrm{T}}}\xspace}
\newcommand{\BC}{\ensuremath{{B_{\mathrm{c}}}}\xspace}
\newcommand{\bbarc}{\ensuremath{{\overline{b}c}}\xspace}
\newcommand{\bbbar}{\ensuremath{{b\overline{b}}}\xspace}
\newcommand{\ccbar}{\ensuremath{{c\overline{c}}}\xspace}
\newcommand{\JPsi}{\ensuremath{{J}/\psi}\xspace}
\newcommand{\bspsiphi}{\ensuremath{B_s \to \JPsi\, \phi}\xspace}
\newcommand{\AFB}{\ensuremath{A_\mathrm{FB}}\xspace}
\newcommand{\EE}{\ensuremath{e^+e^-}\xspace}
\newcommand{\MM}{\ensuremath{\mu^+\mu^-}\xspace}
\newcommand{\TT}{\ensuremath{\tau^+\tau^-}\xspace}
\newcommand{\wangle}{\ensuremath{\sin^{2}\theta_{\mathrm{eff}}^\mathrm{lept}(M^2_\mathrm{Z})}\xspace}
\newcommand{\ttbar}{\ensuremath{{t\overline{t}}}\xspace}
\newcommand{\stat}{\ensuremath{\,\text{(stat.)}}\xspace}
\newcommand{\syst}{\ensuremath{\,\text{(syst.)}}\xspace}

\newcommand{\HGG}{\ensuremath{\mathrm{H}\to\gamma\gamma}}
\newcommand{\gev}{\GeV}
\newcommand{\GAMJET}{\ensuremath{\gamma + \mathrm{jet}}}
\newcommand{\PPTOJETS}{\ensuremath{\mathrm{pp}\to\mathrm{jets}}}
\newcommand{\PPTOGG}{\ensuremath{\mathrm{pp}\to\gamma\gamma}}
\newcommand{\PPTOGAMJET}{\ensuremath{\mathrm{pp}\to\gamma +
\mathrm{jet}
}}
\newcommand{\MH}{\ensuremath{\mathrm{M_{\mathrm{H}}}}}
\newcommand{\RNINE}{\ensuremath{\mathrm{R}_\mathrm{9}}}
\newcommand{\DR}{\ensuremath{\Delta\mathrm{R}}}


\newcommand{\PT}{\ensuremath{p_{\mathrm{T}}}\xspace}
\newcommand{\pt}{\ensuremath{p_{\mathrm{T}}}\xspace}
\newcommand{\ET}{\ensuremath{E_{\mathrm{T}}}\xspace}
\newcommand{\HT}{\ensuremath{H_{\mathrm{T}}}\xspace}
\newcommand{\et}{\ensuremath{E_{\mathrm{T}}}\xspace}
\newcommand{\Em}{\ensuremath{E\!\!\!/}\xspace}
\newcommand{\Pm}{\ensuremath{p\!\!\!/}\xspace}
\newcommand{\PTm}{\ensuremath{{p\!\!\!/}_{\mathrm{T}}}\xspace}
\newcommand{\ETm}{\ensuremath{E_{\mathrm{T}}^{\mathrm{miss}}}\xspace}
\newcommand{\MET}{\ensuremath{E_{\mathrm{T}}^{\mathrm{miss}}}\xspace}
\newcommand{\ETmiss}{\ensuremath{E_{\mathrm{T}}^{\mathrm{miss}}}\xspace}
\newcommand{\VEtmiss}{\ensuremath{{\vec E}_{\mathrm{T}}^{\mathrm{miss}}}\xspace}

%

\newcommand{\ga}{\ensuremath{\gtrsim}}
\newcommand{\la}{\ensuremath{\lesssim}}
\newcommand{\swsq}{\ensuremath{\sin^2\theta_W}\xspace}
\newcommand{\cwsq}{\ensuremath{\cos^2\theta_W}\xspace}
\newcommand{\tanb}{\ensuremath{\tan\beta}\xspace}
\newcommand{\tanbsq}{\ensuremath{\tan^{2}\beta}\xspace}
\newcommand{\sidb}{\ensuremath{\sin 2\beta}\xspace}
\newcommand{\alpS}{\ensuremath{\alpha_S}\xspace}
\newcommand{\alpt}{\ensuremath{\tilde{\alpha}}\xspace}

\newcommand{\QL}{\ensuremath{Q_L}\xspace}
\newcommand{\sQ}{\ensuremath{\tilde{Q}}\xspace}
\newcommand{\sQL}{\ensuremath{\tilde{Q}_L}\xspace}
\newcommand{\ULC}{\ensuremath{U_L^C}\xspace}
\newcommand{\sUC}{\ensuremath{\tilde{U}^C}\xspace}
\newcommand{\sULC}{\ensuremath{\tilde{U}_L^C}\xspace}
\newcommand{\DLC}{\ensuremath{D_L^C}\xspace}
\newcommand{\sDC}{\ensuremath{\tilde{D}^C}\xspace}
\newcommand{\sDLC}{\ensuremath{\tilde{D}_L^C}\xspace}
\newcommand{\LL}{\ensuremath{L_L}\xspace}
\newcommand{\sL}{\ensuremath{\tilde{L}}\xspace}
\newcommand{\sLL}{\ensuremath{\tilde{L}_L}\xspace}
\newcommand{\ELC}{\ensuremath{E_L^C}\xspace}
\newcommand{\sEC}{\ensuremath{\tilde{E}^C}\xspace}
\newcommand{\sELC}{\ensuremath{\tilde{E}_L^C}\xspace}
\newcommand{\sEL}{\ensuremath{\tilde{E}_L}\xspace}
\newcommand{\sER}{\ensuremath{\tilde{E}_R}\xspace}
\newcommand{\sFer}{\ensuremath{\tilde{f}}\xspace}
\newcommand{\sQua}{\ensuremath{\tilde{q}}\xspace}
\newcommand{\sUp}{\ensuremath{\tilde{u}}\xspace}
\newcommand{\suL}{\ensuremath{\tilde{u}_L}\xspace}
\newcommand{\suR}{\ensuremath{\tilde{u}_R}\xspace}
\newcommand{\sDw}{\ensuremath{\tilde{d}}\xspace}
\newcommand{\sdL}{\ensuremath{\tilde{d}_L}\xspace}
\newcommand{\sdR}{\ensuremath{\tilde{d}_R}\xspace}
\newcommand{\sTop}{\ensuremath{\tilde{t}}\xspace}
\newcommand{\stL}{\ensuremath{\tilde{t}_L}\xspace}
\newcommand{\stR}{\ensuremath{\tilde{t}_R}\xspace}
\newcommand{\stone}{\ensuremath{\tilde{t}_1}\xspace}
\newcommand{\sttwo}{\ensuremath{\tilde{t}_2}\xspace}
\newcommand{\sBot}{\ensuremath{\tilde{b}}\xspace}
\newcommand{\sbL}{\ensuremath{\tilde{b}_L}\xspace}
\newcommand{\sbR}{\ensuremath{\tilde{b}_R}\xspace}
\newcommand{\sbone}{\ensuremath{\tilde{b}_1}\xspace}
\newcommand{\sbtwo}{\ensuremath{\tilde{b}_2}\xspace}
\newcommand{\sLep}{\ensuremath{\tilde{l}}\xspace}
\newcommand{\sLepC}{\ensuremath{\tilde{l}^C}\xspace}
\newcommand{\sEl}{\ensuremath{\tilde{e}}\xspace}
\newcommand{\sElC}{\ensuremath{\tilde{e}^C}\xspace}
\newcommand{\seL}{\ensuremath{\tilde{e}_L}\xspace}
\newcommand{\seR}{\ensuremath{\tilde{e}_R}\xspace}
\newcommand{\snL}{\ensuremath{\tilde{\nu}_L}\xspace}
\newcommand{\sMu}{\ensuremath{\tilde{\mu}}\xspace}
\newcommand{\sNu}{\ensuremath{\tilde{\nu}}\xspace}
\newcommand{\sTau}{\ensuremath{\tilde{\tau}}\xspace}
\newcommand{\Glu}{\ensuremath{g}\xspace}
\newcommand{\sGlu}{\ensuremath{\tilde{g}}\xspace}
\newcommand{\Wpm}{\ensuremath{W^{\pm}}\xspace}
\newcommand{\sWpm}{\ensuremath{\tilde{W}^{\pm}}\xspace}
\newcommand{\Wz}{\ensuremath{W^{0}}\xspace}
\newcommand{\sWz}{\ensuremath{\tilde{W}^{0}}\xspace}
\newcommand{\sWino}{\ensuremath{\tilde{W}}\xspace}
\newcommand{\Bz}{\ensuremath{B^{0}}\xspace}
\newcommand{\sBz}{\ensuremath{\tilde{B}^{0}}\xspace}
\newcommand{\sBino}{\ensuremath{\tilde{B}}\xspace}
\newcommand{\Zz}{\ensuremath{Z^{0}}\xspace}
\newcommand{\sZino}{\ensuremath{\tilde{Z}^{0}}\xspace}
\newcommand{\sGam}{\ensuremath{\tilde{\gamma}}\xspace}
\newcommand{\chiz}{\ensuremath{\tilde{\chi}^{0}}\xspace}
\newcommand{\chip}{\ensuremath{\tilde{\chi}^{+}}\xspace}
\newcommand{\chim}{\ensuremath{\tilde{\chi}^{-}}\xspace}
\newcommand{\chipm}{\ensuremath{\tilde{\chi}^{\pm}}\xspace}
\newcommand{\Hone}{\ensuremath{H_{d}}\xspace}
\newcommand{\sHone}{\ensuremath{\tilde{H}_{d}}\xspace}
\newcommand{\Htwo}{\ensuremath{H_{u}}\xspace}
\newcommand{\sHtwo}{\ensuremath{\tilde{H}_{u}}\xspace}
\newcommand{\sHig}{\ensuremath{\tilde{H}}\xspace}
\newcommand{\sHa}{\ensuremath{\tilde{H}_{a}}\xspace}
\newcommand{\sHb}{\ensuremath{\tilde{H}_{b}}\xspace}
\newcommand{\sHpm}{\ensuremath{\tilde{H}^{\pm}}\xspace}
\newcommand{\hz}{\ensuremath{h^{0}}\xspace}
\newcommand{\Hz}{\ensuremath{H^{0}}\xspace}
\newcommand{\Az}{\ensuremath{A^{0}}\xspace}
\newcommand{\Hpm}{\ensuremath{H^{\pm}}\xspace}
\newcommand{\sGra}{\ensuremath{\tilde{G}}\xspace}
\newcommand{\mtil}{\ensuremath{\tilde{m}}\xspace}
\newcommand{\rpv}{\ensuremath{\rlap{\kern.2em/}R}\xspace}
\newcommand{\LLE}{\ensuremath{LL\bar{E}}\xspace}
\newcommand{\LQD}{\ensuremath{LQ\bar{D}}\xspace}
\newcommand{\UDD}{\ensuremath{\overline{UDD}}\xspace}
\newcommand{\Lam}{\ensuremath{\lambda}\xspace}
\newcommand{\Lamp}{\ensuremath{\lambda'}\xspace}
\newcommand{\Lampp}{\ensuremath{\lambda''}\xspace}
\newcommand{\spinbd}[2]{\ensuremath{\bar{#1}_{\dot{#2}}}\xspace}

\newcommand{\MD}{\ensuremath{{M_\mathrm{D}}}\xspace}
\newcommand{\Mpl}{\ensuremath{{M_\mathrm{Pl}}}\xspace}
\newcommand{\Rinv} {\ensuremath{{R}^{-1}}\xspace}

%
%
\hyphenation{en-viron-men-tal}

\cmsNoteHeader{09-005}
\title{Measurement of the Muon Stopping Power in Lead Tungstate}

\address[cern]{CERN}
\author[cern]{The CMS Collaboration}

\date{\today}

\abstract{
A large sample of cosmic ray events collected by the CMS
detector is exploited to measure the specific energy loss of muons in
the lead tungstate (PbWO$_4$) of the electromagnetic calorimeter. The
measurement spans a momentum range from 5 GeV/$c$ to 1 TeV/$c$. The
results are consistent with the expectations over the entire
range. The calorimeter energy scale, set with 120 GeV/$c$ electrons,
is validated down to the sub-GeV region using energy deposits, of
order 100~MeV, associated with low-momentum muons. The muon critical
energy in PbWO$_4$ is measured to be $160^{+5}_{-6} \pm
8$~GeV, in agreement with expectations. This is the first
experimental determination of muon critical energy.}

\hypersetup{%
pdfauthor={A. Benaglia, A. Ghezzi, F. De Guio, A. Martelli,
S. Ragazzi, T. Tabarelli de Fatis, P. Govoni},%
pdftitle={Measurement of the Muon Stopping Power in Lead Tungstate},%
pdfsubject={CMS},%
pdfkeywords={CMS, ECAL, electromagnetic calorimeter, muon stopping power, critical energy}
}
\maketitle 


\section{Introduction}
The electromagnetic calorimeter (ECAL) of the Compact Muon Solenoid
(CMS) experiment~\cite{CMSdetector} has been designed with a stringent
performance goal on energy resolution, driven by the capability to
detect a Higgs boson decaying into two photons. While waiting to
record events from beam collisions at the LHC, its outstanding
characteristics have been exploited to measure the muon stopping power
in lead tungstate ($\textnormal{PbWO}_{4}$) for cosmic ray muon
momenta between 5 and 1000 GeV/$c$.\par

The stopping power for muons in the energy range considered can be
conveniently written as

\begin{equation}
        f(E)=\left\langle -\frac{{\rm d}E}{{\rm d}x} \right\rangle=a(E)+b(E)E\,,
\end{equation}

where $E$ is the total muon energy, $x$ is the thickness of the
traversed material, commonly measured in mass per unit surface, $a(E)$
is the stopping power due to collisions with atomic electrons, and
$b(E)$ is due to radiative processes: bremsstrahlung, direct pair
production, and photonuclear interactions; $a(E)$ and $b(E)$ are
slowly varying functions of $E$ at energies where radiative
contributions are important~\cite{PDG, Striganov}.\par

Numerical values of stopping power and related quantities quoted
throughout this paper are taken from tables in \cite{ANP_Tables}. The
tables \cite[see][]{PDG} are obtained from calculations following
\cite{Striganov} including post-Born corrections from \cite{Ivanov},
which result in an increase of the critical energy by 6 GeV. The
theoretical uncertainties are everywhere smaller than the statistical
precision of the results reported in this paper. The definition of
critical energy as the energy at which the average rates of energy
loss through collision and radiation are equal~\cite{PDG, Striganov}
is adopted in this paper.

The lowest and highest momenta considered in the analysis reported
here correspond to regions  dominated by collision and radiative
processes, respectively. Comparison of the experimental values of the
stopping power with predictions, extends the validity of the ECAL
energy scale, set with 120~GeV/$c$ electron beams, down to energy
deposits below 1~GeV. Analysis of data in the full momentum range
permits a comparison of collision losses with radiative losses, thus
leading to the first experimental determination of muon critical
energy.\par

The experimental setup and the characteristics of the data sample are
described in Section~\ref{Experimental setup and data sample}. Event
selection and measurement of relevant physical quantities are
discussed in Section~\ref{Measurement of physical quantities and event
  selection}. Possible limitations in the measurement of the muon
stopping power are addressed in Section~\ref{Feasibility of the
  measurement of the stopping power}. In Section~\ref{DataAnalysis},
corrections for systematic effects and statistical issues are
addressed, and results are reported.\par

\section{Experimental setup and data sample}
\label{Experimental setup and data sample}
The data set consists of cosmic ray muons traversing the CMS detector
located 100~m underground, collected during the detector commissioning
phase. \par

The central feature of the CMS apparatus is a superconducting solenoid
of 6~m internal diameter.  Within the field volume are the silicon
pixel and strip tracker, the crystal electromagnetic calorimeter and
the brass-scintillator hadron calorimeter (HCAL). Muons are measured
in gas-ionization chambers embedded in the steel return yoke. In
addition to the barrel and endcap detectors, CMS has an extensive
forward calorimetry. The silicon pixel and strip detectors provide
excellent momentum reconstruction, with a $10\, \%$ resolution for
1~TeV/$c$ muons passing close to the nominal interaction vertex.\par

The electromagnetic calorimeter is a hermetic homogeneous scintillator
detector made of PbWO$_4$ crystals. Lead tungstate has a high
density (8.3 g\,cm$^{-3}$), a short radiation length, ($X_0 = 0.89$~cm)
and a small Moliere radius ($R_M = 2.0$~cm). Crystals are organized in a
central barrel of 61\,200 crystals closed by two end-caps of 7324
crystals each. The light is read out by avalanche photodiodes (APD) in
the barrel and by vacuum phototriodes (VPT) in the end-caps. In the
barrel, relevant to the analysis presented in this paper, the
individual crystals have a truncated-pyramid shape with a mean lateral
size in the front section of 2.2~cm and a length of 23~cm, which
corresponds to $25.8$ radiation lengths. The barrel crystals are
organized in a quasi-projective geometry, where their principal axis
is tilted by $3^\circ$ with respect to the nominal interaction
vertex, in both the  azimuthal and polar angle projections.

\begin{figure}[hbtp]
  \begin{center}
    \subfigure[]{\includegraphics[scale=0.35]{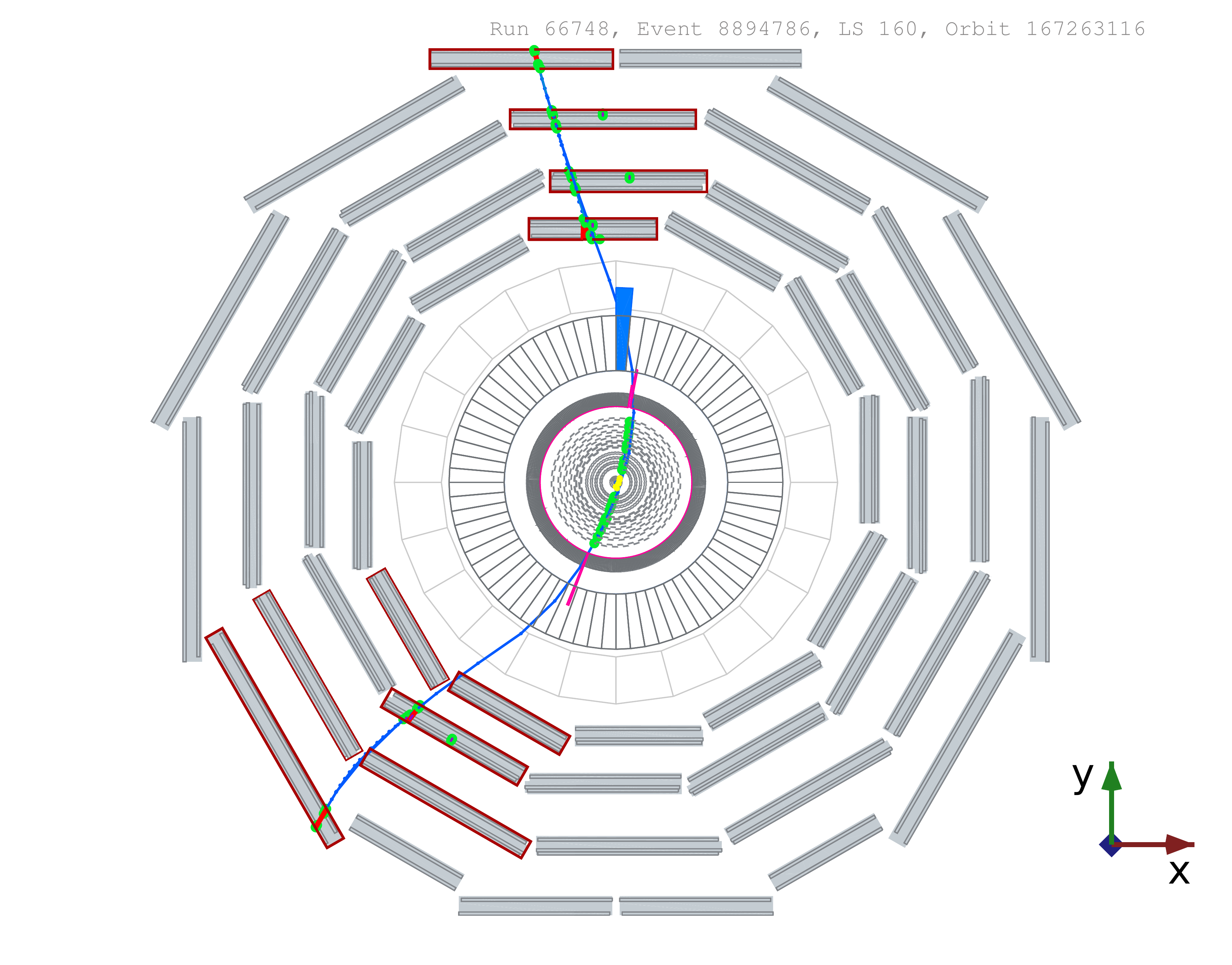}\label{fig:all}}
    \vskip 0.5cm
    \subfigure[]{\includegraphics[scale=0.30]{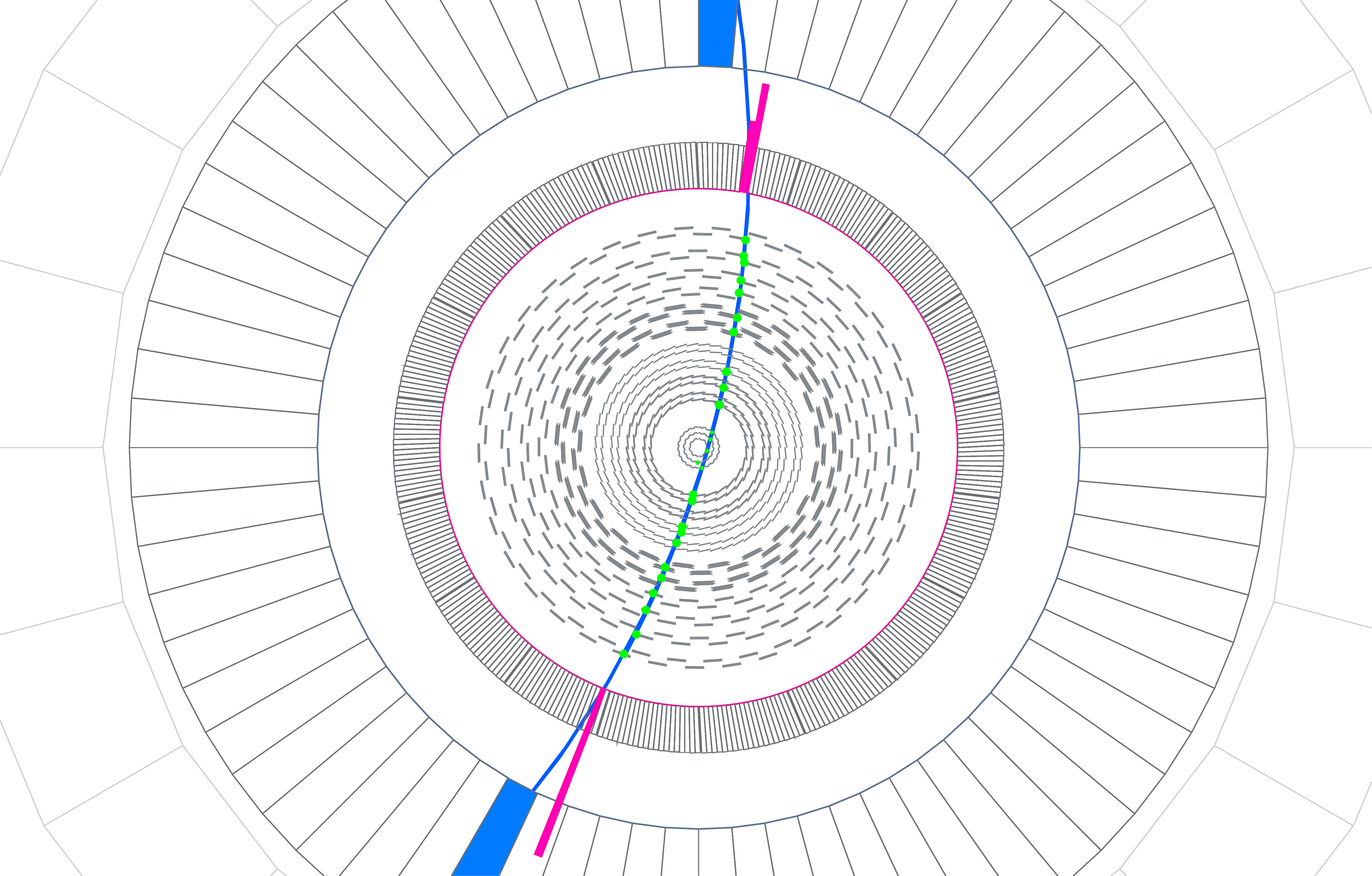}\label{fig:zoom}}
    \vskip 1cm
    \caption {An event display of a cosmic ray muon crossing CMS:
      \subref{fig:all} the full detector and  \subref{fig:zoom} a
      detail of the central region. ECAL hits are in magenta, HCAL hits in
      blue, tracker and muon hits in green. The views represent a
      section of CMS in a plane perpendicular to the beam direction,
      the $x$ axis pointing to the centre of the LHC ring and the $y$
      axis pointing vertically upwards.}
    \label{fig:CMSEcal}
  \end{center}
\end{figure}

The CMS collaboration conducted a month-long data-taking exercise,
known as the Cosmic Run At Four Tesla (CRAFT), during October-November
2008 with the goal of  commissioning the experiment for extended
operation \cite{CFT-09-008}. With all installed detector systems
participating, CMS recorded 270 million cosmic ray triggered events
with the solenoid at its nominal axial field strength of 3.8~T. An
event display of a cosmic ray muon crossing CMS is shown in
Fig.~\ref{fig:CMSEcal}.

During data taking at the LHC, the APDs will be operated at a gain of
50. In order to increase the sensitivity to low-energy deposits and to
reduce the impact of the electronic noise, the APD gain was raised to
200 for several of the commissioning runs. In this condition the noise
per readout channel amounts to about 1~ADC count, which corresponds on
average to 9.3~MeV~\cite{CFT-09-004}.\par

The signal from the APDs, after being amplified and shaped, is
digitized at 40~MHz and ten pulse samples are recorded for each
calorimeter channel. In order to reduce the bandwidth required for
data acquisition, a reduction procedure is applied. A summary of data
reduction algorithms is reported in Section~\ref{Measurement of
  physical quantities and event selection}; for more details the
reader is referred to~\cite{Almeida}.\par

Because of the angular distribution of the cosmic ray muon flux and
the lower sensitivity of the ECAL end-caps to the energy released by a
minimum ionizing particle, only the barrel part of ECAL is used in the
present analysis.

\section{Event measurement and selection}
\label{Measurement of physical quantities and event selection}
The physical variables to be measured for each muon are the momentum,
the path length in ECAL, and the energy lost in ECAL. \par

The muon momentum is measured  from the track fit performed in the
inner tracking system~\cite{CFT-09-014}. \par

The path length in ECAL is estimated by propagating the measured muon
track inside the calorimeter taking into account bending due to the
magnetic field and expected energy loss. The average path length in
ECAL for the selected muon sample is 22.0~cm, with an rms spread of
2.6~cm. 
\par

A dedicated algorithm has been developed to reconstruct the energy
deposited by cosmic ray muons in ECAL. The main differences with
respect to the standard reconstruction algorithm for events collected
during LHC operation are driven by the requirement of good
reconstruction and association efficiency down to low energy and for
energy deposits associated with muons not pointing to the nominal
interaction vertex. Online data reduction is based on zero suppression (ZS),
that is, a readout of channels above the ZS threshold, and on a full readout
of selected regions (SR) of high interest, as identified by a
Selective Readout Processor (SRP), based on the information of the
Level-1 trigger system of ECAL~\cite{Almeida}. For CRAFT runs with APD gain set
to 200, the ZS threshold corresponds to about 20~MeV and the SR
threshold to about 170~MeV. When the condition for SR is met, a matrix
of 3$\times$3 trigger towers is read out. Each tower is comprised of
5$\times$5 crystals, the matrix is centered around the tower with energy deposit
above the SR threshold. The energy reconstruction algorithm identifies
clusters starting from a central crystal (seed) with an energy deposit
of at least 15~ADC counts (139.5~MeV), or from the most energetic of a
pair of adjacent crystals with at least 5~ADC counts (46.5~MeV) each,
the cluster energy is determined as the sum of all channels above
2~ADC counts (18.5~MeV) belonging to a 5$\times$5 matrix of
crystals.\par

Contiguous clusters are merged and the size of the resulting cluster
depends on the impact angle of the muon at the calorimeter
surface. Clusters in ECAL are then associated with muon tracks
according to a geometrical match between the track extrapolation to
the calorimeter surface and the centre of gravity of the energy
deposit. The energy in the cluster is obtained after application of
relative calibration constants to individual channels and of an
absolute energy scale factor. The intercalibration constants are
defined to equalize the response of the individual channels to the
same energy deposits in the corresponding crystals. The absolute
energy scale was determined from data taken with a 120~GeV/$c$
electron beam~\cite{JINST8-10-p10007}. It is defined as the response
to 120~GeV/$c$ electrons striking the centre of a $5 \times 5$ matrix
of crystals in a reference region of ECAL. With these definitions, the
same energy scale is obtained over the entire ECAL up to local
containment effects. According to Monte Carlo simulation, this
containment factor corresponds to $97\,\%$ of the electron energy for
an impact at the point of maximum response of the reference crystal
matrix. Energy deposits due to non-radiating relativistic particles
are fully contained in a $5 \times 5$ crystal matrix, thus the energy
scale defined at the test beam is scaled by 0.97 to measure the ECAL
response to muons. Additional corrections, related to containment
effects due to leakage from the crystals of muon induced secondary
particles, are addressed in Section \ref{DataAnalysis}.\par

Only runs in which the APD gain was set to 200 are considered in the
present analysis. The useful sample of CMS triggers is thus reduced to
$8.8\times10^{7}$. Furthermore, muons are required to cross both the
upper and the lower half of ECAL barrel (see
Section~\ref{DataAnalysis}). The initial sample of events considered
in the present analysis was thus defined by the combined requirements
of a muon at the trigger level~\cite{CFT-09-020}, of a single muon
track reconstructed by the tracker~\cite{CFT-09-014}, and of an energy
deposit associated with the track in both the upper and the lower half
of ECAL barrel. Moreover, analysis is limited to muons with momentum
ranging from 5 to 1000~GeV/$c$. This interval, which is mainly
determined by the size of the event sample, is further divided for
analysis purposes into 20 equal bins in $\log(p)$. For the present
analysis, the muon momentum has to be measured upstream of the energy
release in the calorimeter. Thus only energy deposits in the bottom
half of ECAL are used. The following selections are then applied to
the initial sample: the uncertainty on muon momentum must be smaller
than the bin width; the ratio of the energy deposit in the calorimeter
to muon momentum must be less than 1 within the experimental accuracy;
the angle between the muon track at the ECAL surface and the
corresponding crystal axis must be smaller than 0.5~radians. These
criteria reduce the sample by a factor 1.7, the largest reduction
coming from the angle selection. Moreover, as will be clarified in
Section~\ref{DataAnalysis}, events with energy deposits above 500~MeV
in the upper part of the ECAL barrel are vetoed. This last selection
further reduces the sample by a factor 1.15. The final sample consists
of $2.5 \times 10^{5}$ muons.\par

\begin{figure}[hbtp]
  \begin{center}
    \subfigure{\includegraphics[scale=0.39, angle=0]{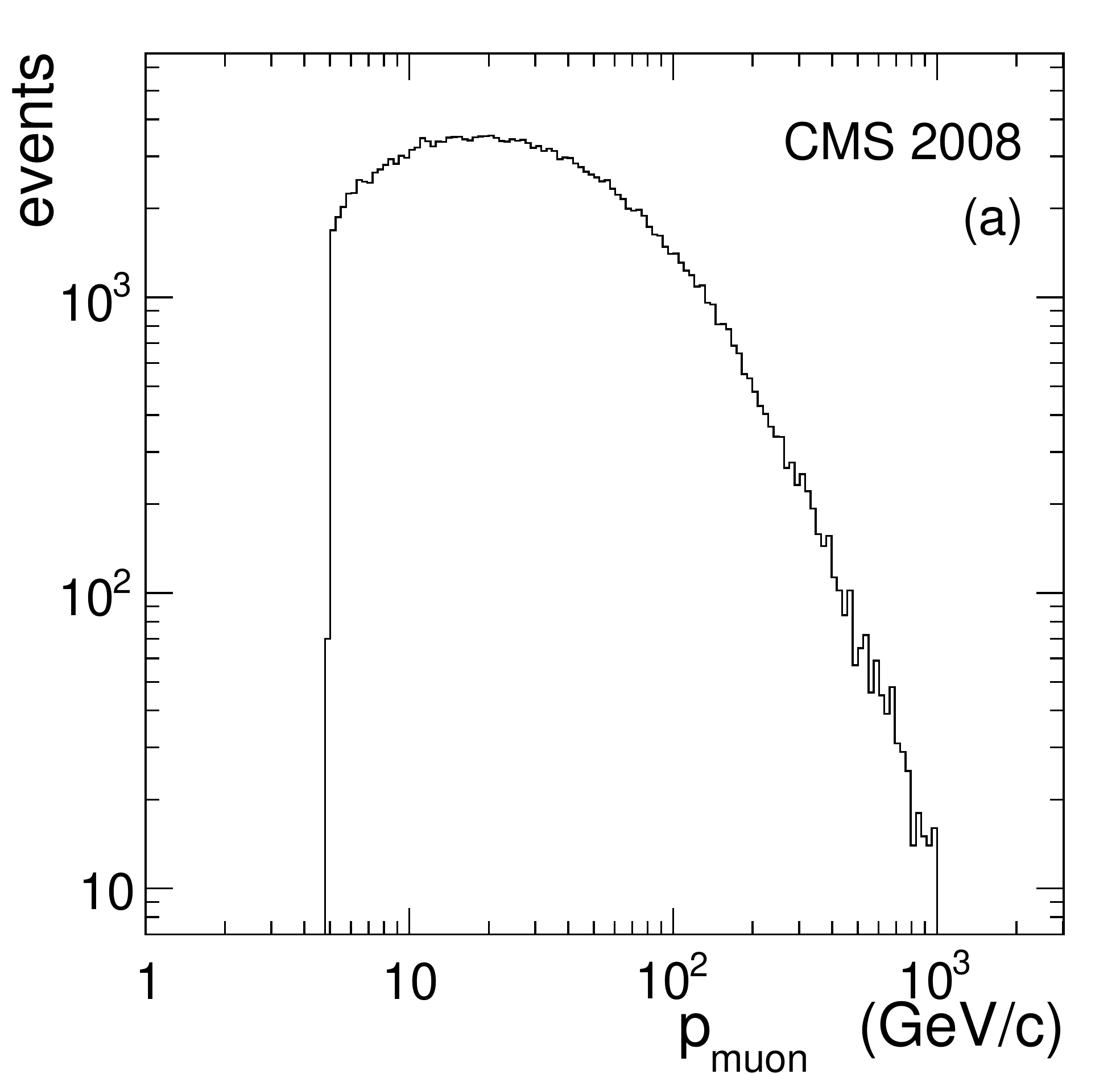}\label{fig:spectraMu}} 
    \subfigure{\includegraphics[scale=0.39, angle=0]{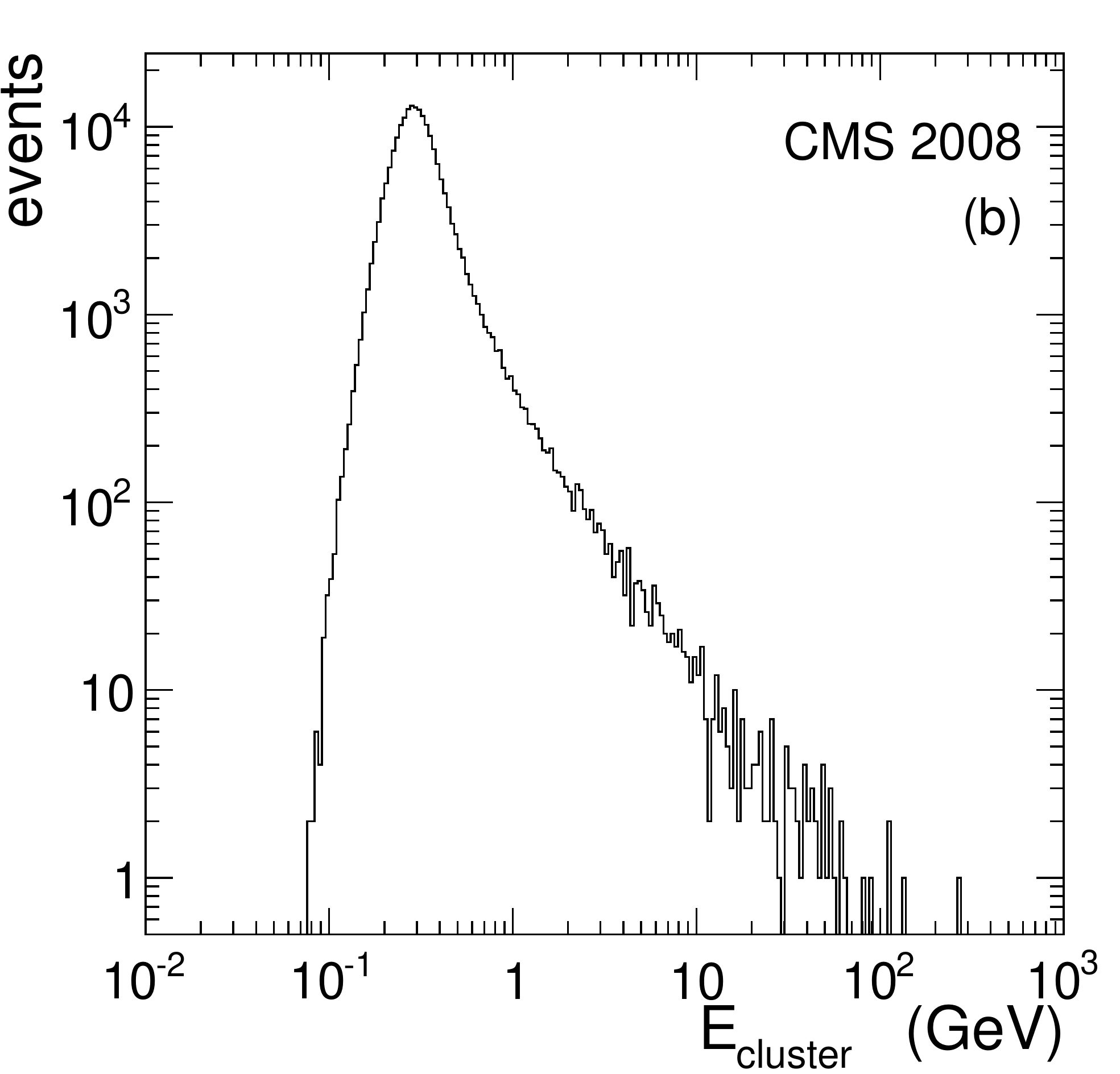}\label{fig:spectraEcal}} 
    \hspace{1cm}
    \caption{ \subref{fig:spectraMu} Momentum spectrum of the muons
      passing the selections;  \subref{fig:spectraEcal} spectrum of
      the reconstructed energy in the lower ECAL hemisphere. A
      logarithmic binning is used.}
    \label{fig:spectra}
  \end{center}
\end{figure}

The measured muon spectrum, after the selections, is shown in
Fig.~\ref{fig:spectra}\subref{fig:spectraMu}. The spectrum of energy
deposited in crystals is shown in
Fig.~\ref{fig:spectra}\subref{fig:spectraEcal}.

\begin{figure}[hbtp]
  \begin{center}
    \subfigure{\includegraphics[scale=0.39, angle=0]{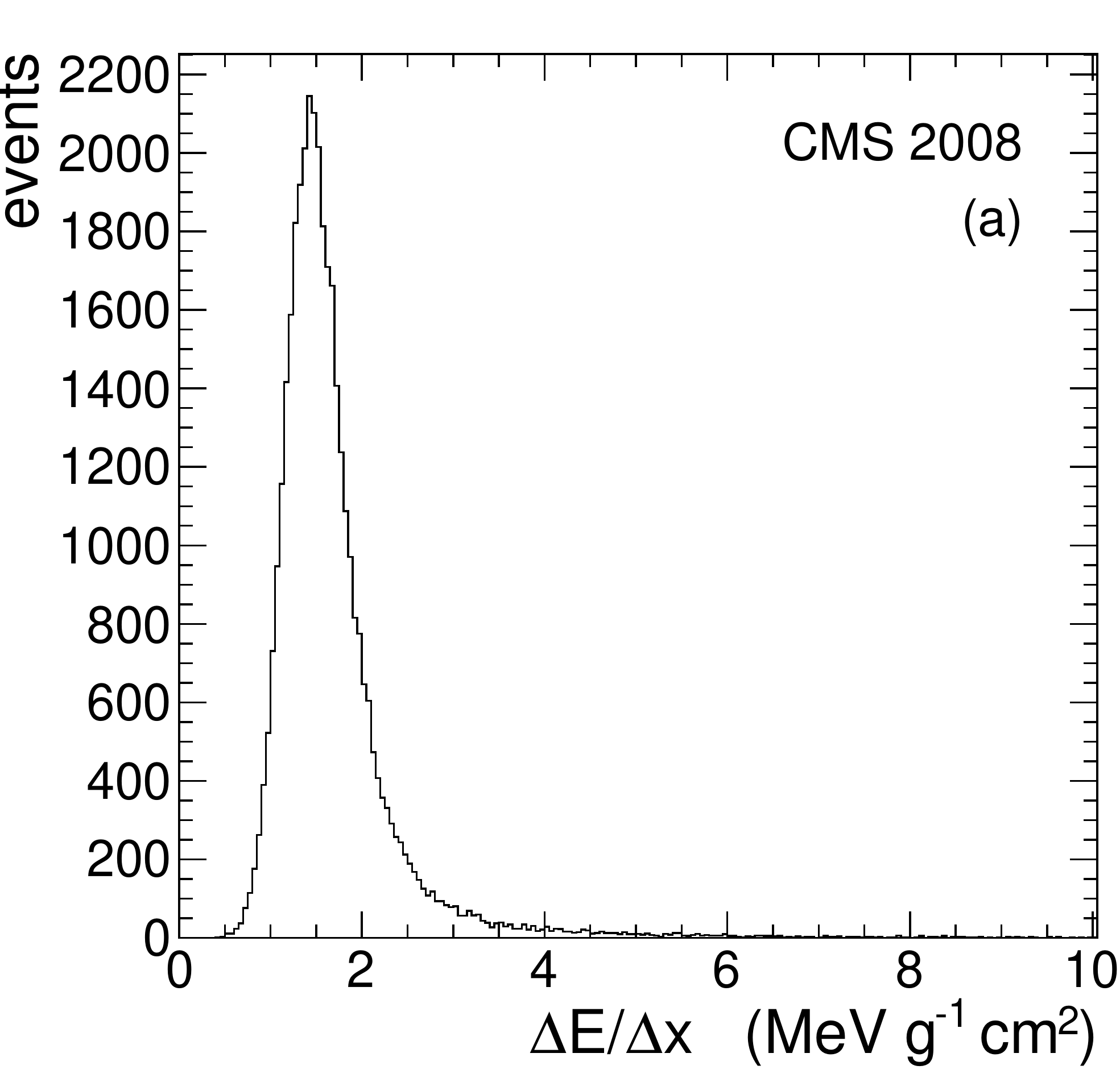}\label{fig:dEodxLow}} 
    \subfigure{\includegraphics[scale=0.39, angle=0]{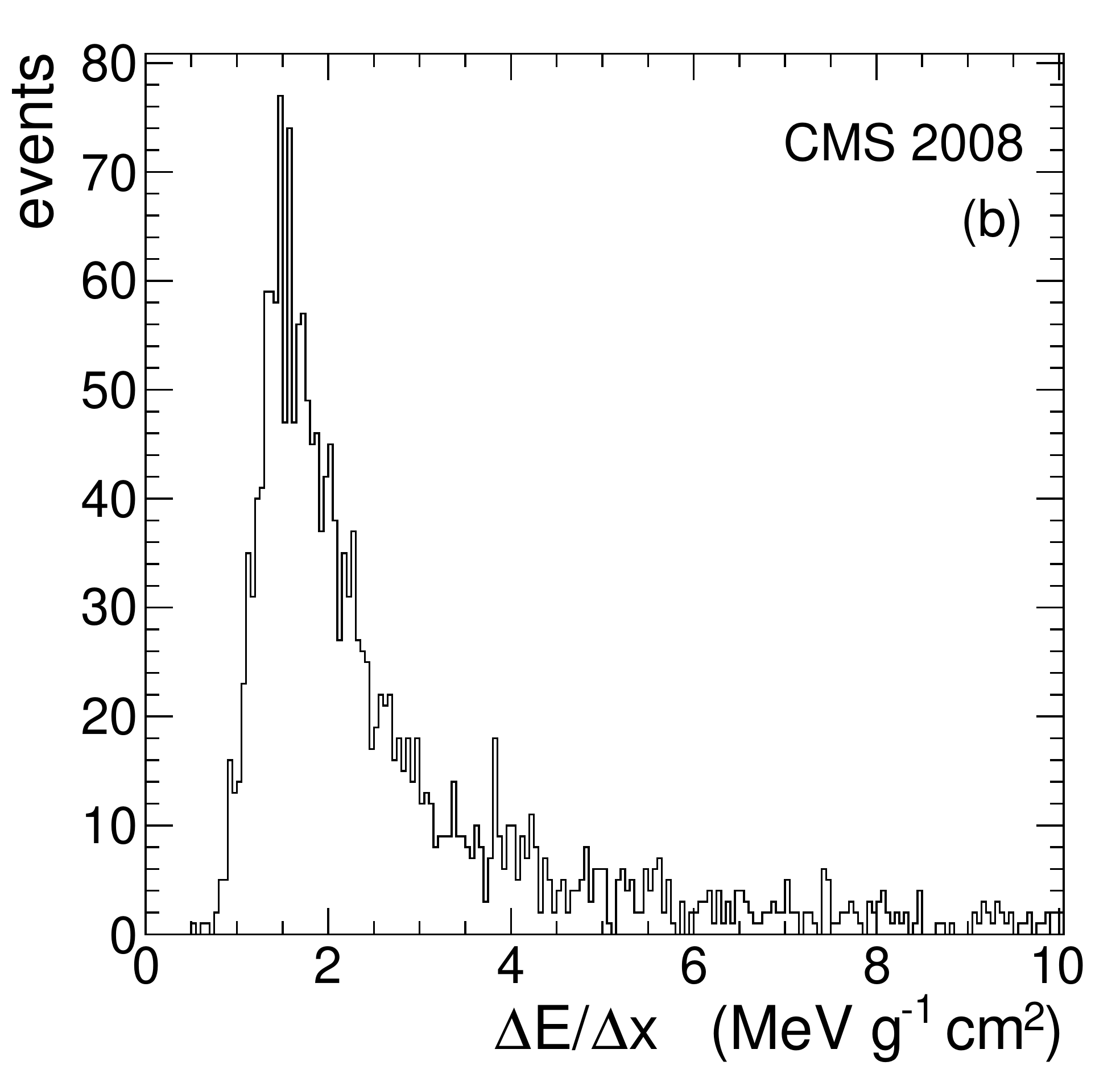}\label{fig:dEodxHigh}} 
    \hspace{1cm}
    \caption{Measured distributions of $\Delta E / \Delta x$ in ECAL;
      \subref{fig:dEodxLow} for muon momenta below 10~GeV/$c$;
      \subref{fig:dEodxHigh} for muon momenta above 300~GeV/$c$;
      the fraction of events with $\Delta E/\Delta x >
      10$~MeV\,g$^{-1}$\,cm$^2$ is $1.3\times10^{-3}$ and
      $8\times10^{-2}$ in \subref{fig:dEodxLow} and
      \subref{fig:dEodxHigh} respectively.}
    \label{fig:dEdxDist}
  \end{center}
\end{figure}

Figure~\ref{fig:dEdxDist} displays the distributions of $\Delta E /
\Delta x$, the measured cluster energy divided by the path length in
ECAL, for muons with momentum below 10~GeV/$c$, where collision losses
dominate \subref{fig:dEodxLow}, and for muons with momentum above
300~GeV/$c$, where sizable radiation losses are expected
\subref{fig:dEodxHigh}.

\section{Measurement of the stopping power}
\label{Feasibility of the measurement of the stopping power}
In making a calorimetric measurement of stopping power, the choice of
optimal thickness of the sampling material must be the result of a
compromise. On the one hand, the thickness should be small enough for
the variation of primary particle energy to be negligible; on the
other, the thickness should be large enough to allow a precise
measurement of the energy released.  Furthermore, the correction
arising from energy leaking from the measurement volume should be kept
small. The extent to which the CMS ECAL crystals meet these
requirements is discussed in~\ref{$dE/dx$ approximation}
and~\ref{Containment effects}.

\subsection{\texorpdfstring{${\rm d}E/{\rm d}x$}{{\rm d}E/{\rm d}x} approximation}
\label{$dE/dx$ approximation}
The stopping power is experimentally determined by the measurement of
the energy $\Delta E$ lost by a muon when traversing a known thickness
$\Delta x$, which must be small in order that the approximation
$f(E)=-{\rm d}E/{\rm d}x=\Delta E/ \Delta x$ holds. Quantitatively
$\Delta f/f$, where $\Delta f=f(E)-f(E-\Delta E)$, has to be smaller
than the desired relative precision on the measurement of $f$. Since

\begin{equation}
        \Delta f \sim \frac{{\rm d}f}{{\rm d}E} \frac{{\rm d}E}{{\rm d}x} \Delta x
\end{equation}

then

\begin{equation}
        \frac{\Delta f}{f} \sim \frac{{\rm d}f}{{\rm d}E} \Delta x \,.
\end{equation}

In the present case, $\Delta x$ is typically 180~g/cm$^2$ (about
$25\,X_0$ in $\textnormal{PbWO}_4$). The derivative of the stopping
power with respect to $E$ can be written as a sum of the contributions
of collision and radiation processes, namely

\begin{equation}
        \frac{\Delta f}{f} = \left [\left ( \frac{{\rm d}f}{{\rm d}E} \right ) _{coll} + \left ( \frac{{\rm d}f}{{\rm d}E} \right ) _{rad}\right ]\Delta x \,.
\end{equation}

For the scope of this analysis, $b(E)$ in Eq.~1 can be assumed to be
constant, with a value equal to that at 1~TeV. Then $({\rm d}f/{\rm
  d}E)_{rad} = b = 1.4\times 10^{-5} \,
\textnormal{cm}^2/\textnormal{g}$~\cite{ANP_Tables}.

This corresponds to a muon radiating 1/400 of the energy when
traversing $\Delta x = 180~\textnormal{g/cm}^2$. Its average
contribution to $\Delta f / f$ is then 1/400, which is negligible when
compared to the precision of the measurement.\par

The component of the derivative due to collisions is a monotonically
decreasing function of $E$ in the energy range of the present
measurement; its expected value at the lowest momentum (5~GeV/$c$) is
$2.6\times 10^{-5}~\textnormal{cm}^2/\textnormal{g}$~\cite{ANP_Tables}.
Thus $\Delta f / f$ is everywhere small when compared with other sources
of uncertainty.\par

The above arguments apply not only to muons losing an energy equal to the
expectation value, but also to the case of large fluctuations occurring in radiative processes.
In fact $f(E)$ is essentially a linear function of $E$ in this energy range.\par

\subsection{Containment effects}
\label{Containment effects}
Energy lost by a muon is transferred to energy of secondaries. Since
secondaries and their daughters may travel a significant distance, the
energy lost by a muon traversing a given volume is in general
different from the energy deposited in that same volume. The energy
$\Delta E$ considered in the previous section is the energy lost by a
muon along a given path, while the first output of the measurement
procedure is the energy deposited in a given volume of the
calorimeter. Taking as a reference volume a cylinder with its axis
along the muon direction, energy flow across the lateral surface can
be made negligible by choosing a sufficiently large radius. On the
other hand there is always a non-negligible flow of energy carried by
secondaries entering and leaving the two ends of the cylinder, since
secondaries are produced along the entire muon path. In a homogeneous
material an equilibrium between the energy flowing in through the
front surface and the energy flowing out of the rear surface is
approximately reached when the front surface is preceded by at least
an amount of material corresponding to the energy-weighted average
range of secondaries. In fact, considering a muon entering a dense
homogeneous material from vacuum, the average energy deposition per
unit thickness is given by the convolution product of a step function
equal to the muon stopping power (muon passing from vacuum to
homogeneous material) with the function $S(t)$ describing the average
fractional energy deposition density as a function of depth $t$ for
secondaries (i.e. the shower depth-profile for high-energy electrons
and photons). This convolution product at depth $x$ in the material is
equal to the muon stopping power times the definite integral from $0$
to $x$ of $S(t)$. Considering, as an illustration, the non-physical
case of a muon whose only interaction with matter is photon radiation,
at a depth in the dense material at which an electromagnetic shower
would have deposited 90\,\% of its energy, the ratio between the
average energy deposition density and the stopping power would be
0.9. Therefore, considering the spectra of secondaries produced by
muons, it is found that the equilibrium condition is approached for
$\delta$-electrons after the muon has traversed a thickness of about
10~g/cm$^2$, while for radiated photons it is approached after about
20 radiation lengths.\par

In the present case, the situation is well described as a homogeneous
region of PbWO$_4$ preceded by order of 10~g/cm$^2$ of nearby
material. Because of the presence of a strong magnetic field, material
separated by more than a few tens of centimeters from the crystals is
less effective in producing secondaries flowing into the measurement
volume. Thus the crystals diametrically opposite to those under study
(corresponding to a separation of approximately 2.6~m) have little
effect on the measured energy loss. Therefore, net containment
corrections are expected to be small or negligible for collision
losses, but sizable for radiative losses. A quantitative estimate of
these corrections has been obtained by means of a dedicated simulation
based on the Geant4 package~\cite{Geant4}, as detailed in Section
\ref{DataAnalysis}.


\section{Data analysis and experimental results}
\label{DataAnalysis}
This section first addresses corrections to systematic effects;
instrumental effects are considered in~\ref{Instrumental effects}; the
physics effects identified in Section~\ref{Feasibility of the
  measurement of the stopping power} are considered
in~\ref{Containment corrections}. Statistical analysis of data and
experimental results are then discussed in~\ref{Results}.

\subsection{Instrumental effects}
\label{Instrumental effects}
The relevant instrumental effects are related to the online data
reduction and to the energy reconstruction procedure described in
Section~\ref{Measurement of physical quantities and event
  selection}.\par

The presence of thresholds introduces a bias in energy reconstruction:
noise fluctuations above threshold contribute a positive bias, while
energy deposits below threshold contribute a negative bias.\par

When the muon direction is close to the crystal axis, the energy
deposited in ECAL is in most cases above the SR threshold (90\,\% of
events with the angle between muon and crystal axis smaller than
0.1~radians). When this condition is met, the bias arises mainly from
effects associated with the clustering threshold and the noise
spectrum, signals below threshold making a negligible
contribution. The probability of noise fluctuation above threshold is
measured in a dedicated analysis of the noise spectrum. A 14.7~MeV
bias is estimated for muons at angles smaller than 0.1~radians. At
larger muon angles to the crystal axis, the average energy deposit per
crystal decreases. Therefore conditions for ZS read-out occur more
frequently, with the higher ZS threshold reducing the noise
bias. Moreover, the probability that the energy deposited in a single
crystal is below clustering threshold increases, thus contributing a
negative bias. Bias is then expected to decrease with increasing
angles, but its estimate depends on several contributions, and is less
reliable than the estimate at small angles. Both a Monte Carlo
simulation and the analysis of crystal multiplicity versus angle in
experimental data indicate a constant positive bias (plateau) at small
angles, followed by a monotonic decrease at larger angles.

\begin{figure}[hbtp]
\begin{center}
\includegraphics[scale=0.39, angle=90]{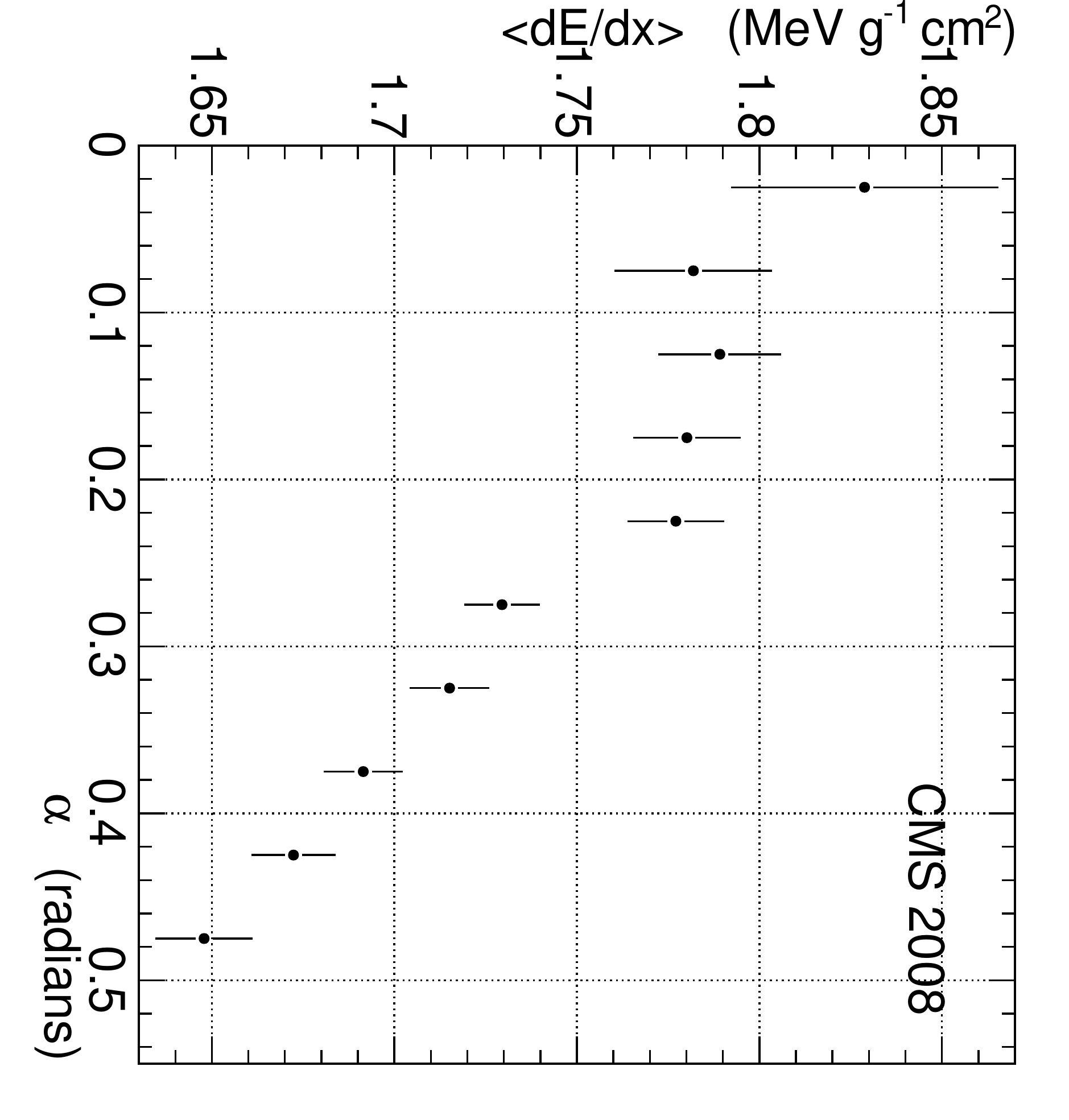} 
\caption{Dependence of the raw $\langle {\rm d}E/{\rm d}x \rangle$ on the
  angle $\alpha$ between the muon direction and the crystal axis, for
  muon momentum between 5 and 10~GeV/$c$. Vertical bars represent
  statistical errors.}
\label{fig:Residuals_vsAngle}
\end{center}
\end{figure}

Bias dependence on angle for non-radiating muons has been further
studied by analyzing the dependence of the raw ${\rm d}E/{\rm d}x$ on angle for
muons below 10 GeV/$c$ in experimental data. The result, shown in
Fig.~\ref{fig:Residuals_vsAngle}, is consistent with expectations,
although it is not inconsistent with a wider plateau, extending up to
0.2~radians, or with a linear dependence of bias on angle with no
plateau. A bias correction, dependent on angle, is thus applied to the
estimated collision component of the stopping power, by applying the
following procedure. At small angles, the bias to be subtracted is
14.7~MeV. At larger angles, the bias to be subtracted is estimated
from a fit to the data shown in Fig.~\ref{fig:Residuals_vsAngle},
assuming a plateau followed by a linear decrease with angle. The
central value of the bias estimate is obtained assuming that the
plateau extends up to 0.1~radians, the hypotheses of no plateau and of
a plateau up to 0.2~radians are taken as extreme cases in the
evaluation of the systematic uncertainty of bias subtraction. The
uncertainty in  the evaluation of the 14.7~MeV bias and the
statistical uncertainty of the fit also contribute to the systematic
uncertainty of the bias correction. The overall systematic uncertainty
is estimated to be 3.5~MeV, which corresponds to about 1.2\,\% of the
average energy deposited by a low-momentum muon.\par

The bias is estimated to be negligible in the presence of the larger
energy deposits associated with radiative processes; it is in any case
negligibly small compared to the systematic effects discussed in
Section \ref{Containment corrections}. Therefore, no bias correction
is applied to the expected contribution to ${\rm d}E/{\rm d}x$ from radiative
processes over the entire muon momentum spectrum.\par

\subsection{Containment corrections}
\label{Containment corrections}
Energy leakage and the effects of upstream and downstream materials
are quantitatively different for collisions and radiative
processes. Therefore, in order to make an experimental determination
of the muon critical energy, losses and compensation for radiative and
collision processes are treated separately.\par

To estimate the difference between energy lost and energy deposited in
ECAL, a model based on the Geant4 simulation toolkit, with a simplified
description of the detector and materials surrounding it, was
developed. The use of a simplified simulation instead of the full
simulation available in CMS was driven by the high statistics needed
for these studies and for those reported in Section~\ref{Results}. The results
of the simulations were also compared with dedicated analytical
calculations.\par

Because of the effect of the intense magnetic field on the
trajectories of charged particles, material lying more than a few tens
of centimeters away from the crystals under study may be neglected. In
particular, the upper half of the ECAL has been ignored in the
simulation. The validity of this approximation has been reinforced by
selecting muons that produce a signal of less than 500~MeV in the
upper part of the ECAL, thus removing most of those emitting a hard
photon.
As a further check, the energy deposited in the lower half of ECAL by
muons releasing less than 500~MeV in the upper ECAL hemisphere (the
selected sample) has been compared with the complementary sample and
agreement was found within the precision of the comparison. In fact,
in the muon momentum region between 145 and 1000~GeV/$c$, where
radiative  losses are relevant, the weighted average of the ratio of
the ${\rm d}E/{\rm d}x$ measured in the two samples in seven momentum
bins is $0.95 \pm 0.09$. This result gives no indication of an effect
of the upper part of ECAL on the energy deposited in the lower part,
thus supporting the hypothesis that far material can be neglected.
However, due to the modest statistical precision of this test, it was
conservatively decided to activate the veto in the event
selection.\par

The low-momentum region is of major interest for the present analysis,
since the large sample size and the small energy loss fluctuations
allow precise results to be obtained. According to the simulation, the
energy carried by secondaries flowing out of the rear detector surface
is, on average, 3\,\% of the energy lost in the crystals for a muon
with momentum 15~GeV/$c$. This represents an upper limit to
containment corrections in the low-momentum region, where collision
losses dominate. In fact, as already discussed in
Sect.~\ref{Containment effects}, the rear leakage can be compensated
by secondaries produced in upstream material entering the front
detector surface. A comparison of the deposits in the upper and lower
hemispheres of the ECAL barrel, for muons with momentum between 5 and
10~GeV/$c$, shows no difference in ${\rm d}E/{\rm d}x$, with a
sensitivity better than 1\,\%~\cite{CFT-09-004}. Given the similarity
of the results, in spite of the asymmetric distribution of the outer
and inner material around ECAL, the net correction has been neglected,
and 1\,\% systematic uncertainty on the null correction for collision
losses has been assumed. \par

Estimated corrections for energy containment in radiative processes
have been derived from dedicated simulations of two extreme cases:
with no material in front of ECAL and with the whole material budget
of the tracker concentrated just in front of ECAL. The former case
represents the upper limit to the containment correction, where only
rear losses are considered; the latter case gives a lower limit to the
correction, as it maximally overestimates the energy flow through the
front face due to upstream material. In both cases the effects of the
upper ECAL hemisphere, used as a veto in the analysis, were ignored.
The mean value and half of the difference between the
results of the two simulations are the estimate of the corrections for
the energy containment in radiative processes and of the associated
systematic error. The correction for energy leakage in radiative
processes is maximal at 1~TeV/$c$: at this muon momentum it
corresponds to $(28\pm 5)\,\%$ of the average energy lost, while at
170~GeV/$c$ it reduces to $(14.5\pm 2.5)\,\%$, where the error arises
from the systematic uncertainty mentioned above. The uncertainty is
small compared to the statistical uncertainty in the measurement of
the stopping power in the corresponding region. Given the size of
statistical uncertainty and the contribution from other sources of
systematic errors, the precision of the correction obtained with the
simplified simulation is sufficient over the entire momentum
range. The correction for non-containment of radiated secondaries is
applied over the whole momentum spectrum to the expected contribution
of radiative processes to stopping power.

\subsection{Critical energy, energy scale, Cherenkov contribution}
\label{Results}
Several possible sources of systematic error have been taken into account in making a comparison
between the energy scale of the ECAL set with a 120 GeV/$c$ electron beam and the scale resulting
from the measurement of muon stopping power. \par

Four out of 36 supermodules were calibrated with 120 GeV/$c$ electrons. The energy scale was
transferred to all supermodules by means of the cosmic ray intercalibration discussed in~\cite{JINST8-10-p10007}.
This procedure resulted in a typical single-channel uncertainty of 1.5\%, which is dominated by statistics,
and an overall scale uncertainty whose contribution to the final precision of the present measurement is negligible.\par

The precise variation in the gain of each APD, when going from a nominal gain of 50
to a nominal gain of 200, has been measured by means of the laser monitoring system
described in~\cite{CMSdetector}. The resulting uncertainty on the average energy scale is negligible.\par 

The overall variation of the response to incident electrons with temperature has been
measured in test beam to be $-(3.8\pm 0.4)\% ^\circ$C$^{-1}$. In CMS, the temperature
is monitored with a $0.01 ^\circ$C precision, and the cooling system
has been demonstrated to maintain the temperature stable
to $18\pm 0.1 ^\circ$C~\cite{CMSdetector, CFT-09-004},
thus uncertainties on temperature corrections are negligible.\par

A test of reproducibility of calibration constants was performed by exposing the same
module to test-beam electrons on two occasions, separated by 30 days. 
This resulted in no measurable offset on the energy scale~\cite{JINST8-10-p10007}.\par

A different pulse height reconstruction procedure is required for cosmic-ray data,
which are asynchronous with respect to detector clock, compared to collision data,
which are synchronous, or to test-beam data, which are also asynchronous but were
recorded together with accurate timing information from trigger. This has been estimated
to introduce a systematic uncertainty of 1\% in the energy scale of cosmic ray data with respect to test beam.\par

\begin{figure}[hbtp]
  \begin{center}
     \subfigure{\includegraphics[scale=0.45, angle=0]{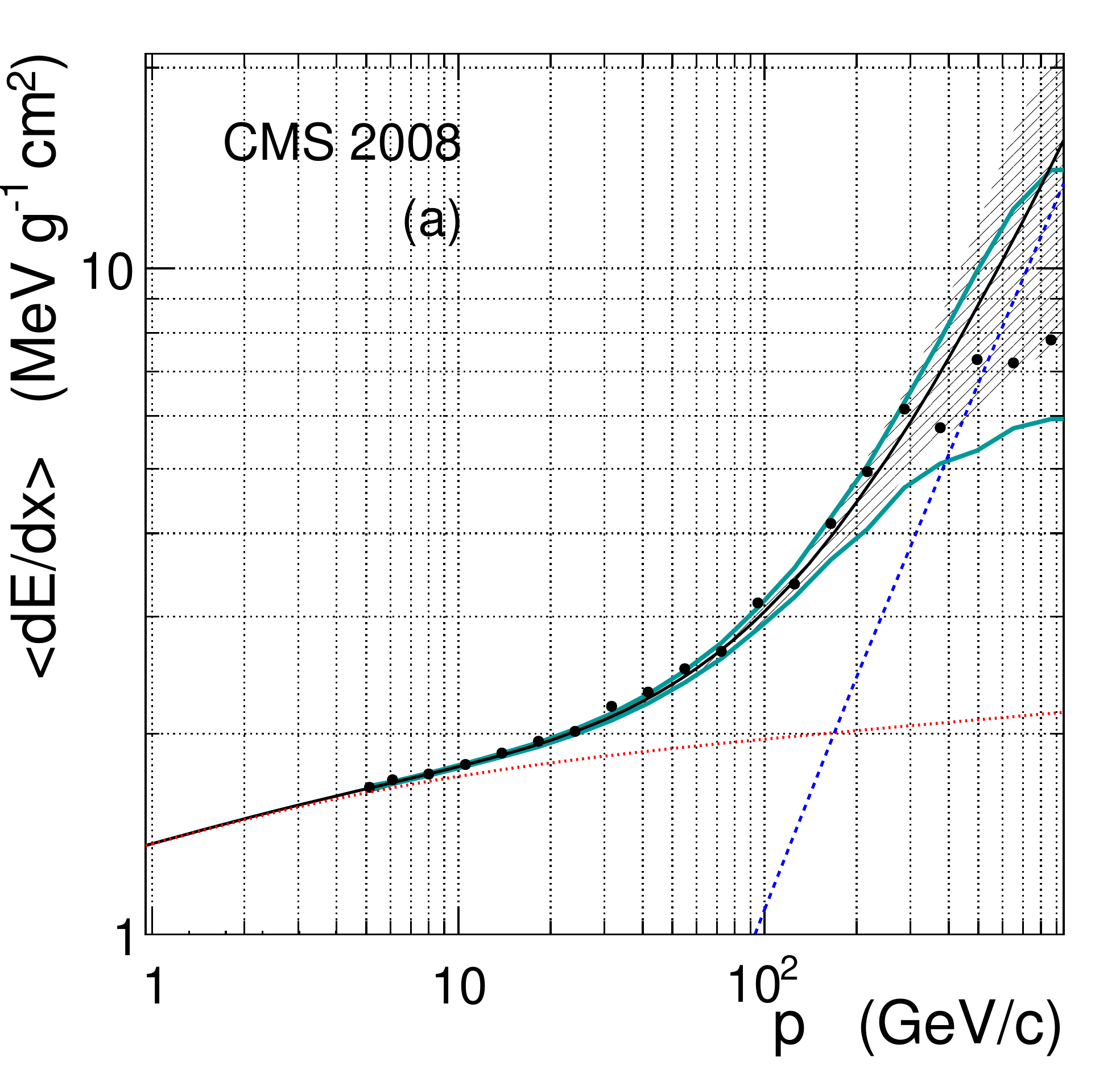}\label{fig:dEodx}}
     \subfigure{\includegraphics[scale=0.45, angle=0]{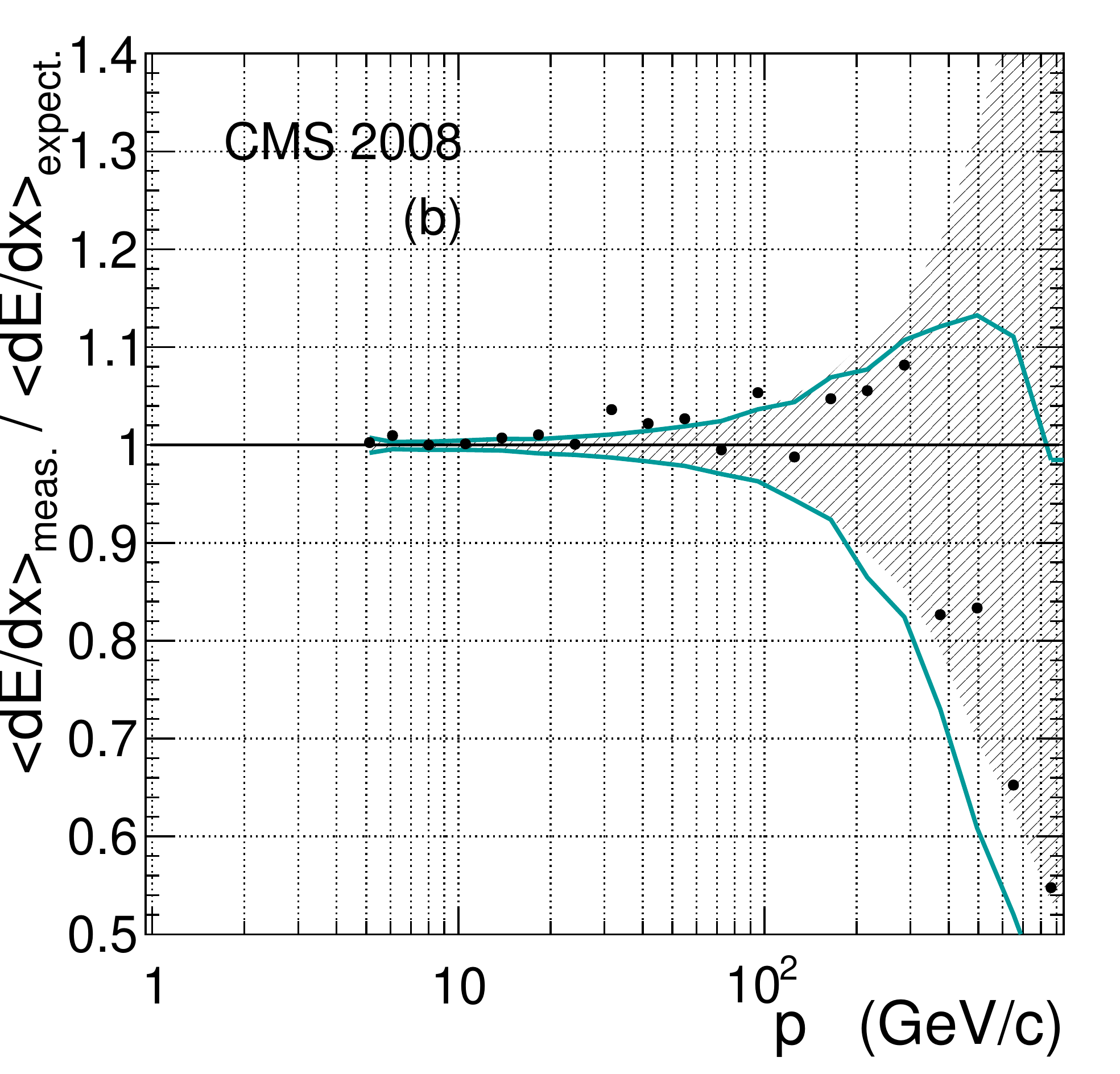}\label{fig:residuals}}
     \caption{\subref{fig:dEodx} Muon stopping power measured in
       PbWO$_4$ (dots) as a function of muon momentum compared to
       expectations \cite{ANP_Tables} (continuous black line). The expected
       contributions from collision and radiative processes are
       plotted as well (red dotted line and blue dashed line
       respectively). \subref{fig:residuals} Ratio of the measured and
       the expected values of the muon stopping power, as a function
       of muon momentum. As discussed in the text, in both figures, the shaded grey
       area indicates the expected 68\,\% probability central
       interval, while the continuous cyan curves delimit the minimum
       interval containing 68\,\% of the expected
       results. ``Experimental errors'' are not shown for reasons
       discussed in the text.}
    \label{fig:dEoverdx_final}
\end{center}
\end{figure}

In Fig.~\ref{fig:dEoverdx_final}\subref{fig:dEodx}, the specific muon
energy loss resulting after corrections is compared to expectations as
a function of the muon momentum~\cite{ANP_Tables}.
Fig.\,\ref{fig:dEoverdx_final}\subref{fig:residuals}
shows the ratio of experimental measurements to expected values. Two
regions are indicated in Fig.~\ref{fig:dEoverdx_final}: the
expected 68\,\%~probability central interval (grey shaded area), and
the minimum interval corresponding to 68\,\%~probability (region
delimited by continuous curves). The reasons for considering both intervals
and the procedure adopted for their estimate are summarized below.\par

The characteristics of radiative losses, which correspond to rare
processes with high energy releases, generate probability distribution
functions (pdf) of energy loss for single events with long tails at
high energy, that is: highly positively skewed distributions. In this
condition, convergence to a normal distribution for the average ${\rm
  d}E/{\rm d}x$ of a sample of $N$ events is slow. The skewness of the
single event distribution increases with increasing muon momentum,
owing to the increasing contribution of radiative losses with
momentum. The number of events per bin in the selected data sample
rapidly decreases with momentum for $p>50$~GeV/$c$
(Fig.~\ref{fig:spectra}\subref{fig:spectraMu}). It is thus expected
that, due to the combined effect of increasing skewness of single
event distribution and decreasing event statistics with increasing
momentum, in the higher momentum bins the pdf of the sample mean of
${\rm d}E/{\rm d}x$ cannot be approximated by a Gaussian. Moreover,
when this condition occurs, sample mean and sample variance are highly
correlated. It follows that the widely used ``experimental error'',
namely the sample rms divided by the square root of the sample
population, is not a good measure for the consistency of the sample
mean with the expected value. Therefore the uncertainty of the result
has been estimated, consistently over the entire momentum range
covered by the measurement, by means of a numerical technique based on
the simulation of the expected results with Geant4. Ten thousand
identical experiments, each with the same statistics per momentum bin
as the actual data sample, were simulated. The expected pdf of the
mean is then obtained for each bin as the distribution of the mean
values of the stopping power from the different experiments. For each
pdf two 68\,\%  probability intervals for the expected result were
derived: the central interval, obtained by discarding 16\,\% of the
results on each tail of the pdf, and the interval of minimum width
containing 68\,\% of the results; this last interval always contains
the most probable value. For Gaussian distributions these intervals
coincide. In Fig.~\ref{fig:dEoverdx_final}, for the event samples
corresponding to the bins above 100 GeV/$c$, the differences between
the two intervals become increasingly significant. At lower momenta,
where the two intervals tend to coincide, they also correspond to the
conventional estimate of the experimental error.

For the high momentum bins, the increasing probability of large
fluctuations in radiative energy losses, combined with the decreasing
size of the event samples, cause the most probable values to be
systematically lower than the expectation values. This trend is
particularly marked in the highest momentum bin of
Fig.~\ref{fig:dEoverdx_final}, where the expectation value lies
outside the 68\% probability interval.\par

The curve

$$ \left({\rm d}E/{\rm d}x\right)_{meas} = \alpha\left[\left(\frac{{\rm d}E}{{\rm d}x}\right)_{coll} + \beta\times \left(\frac{{\rm d}E}{{\rm d}x}\right)_{rad}\right]\,,$$

where $coll$ and $rad$ label the predicted energy losses in
$\textnormal{PbWO}_{4}$ due to collisions with atomic electrons and
radiative processes respectively~\cite{ANP_Tables}, is fitted to
experimental stopping power data using a binned maximum likelihood and
the pdf described above. The parameters $\alpha$ and $\beta$ account
for the overall normalization of the energy scale and for the relative
normalization of radiation and collision losses. With the adopted
parameterization the $\beta$ parameter, from which the critical energy
is measured, is independent of the overall energy scale. The fit
results in:

\begin{align*}
\alpha &= 1.004^{+0.002}_{-0.003}\,({\rm stat.})\pm 0.019\,({\rm syst.})\\
\beta  &= 1.07^{+0.05}_{-0.04}\,({\rm stat.})\pm 0.6\,({\rm syst.}).
\end{align*}

Adding statistical and systematic contributions in quadrature, it may
be concluded from the above results that the energy scale is
consistent with expectations within a systematic uncertainty of
1.6\,\%, with a 1.2\,\% contribution from the uncertainty on the
energy scale dependence on the angle and on the clustering, and a
1.0\,\% contribution from uncertainty in containment corrections. This
result is mainly determined by the precision of the measurements in
the momentum region below 20~GeV/$c$, where radiation losses are
marginal. The typical energy released in this region is about
300~MeV. Thus the energy scale set with 120~GeV/$c$ electrons still
holds in the sub-GeV region. This, assuming a linear response of the
detector to the collected light, implies that the relative
contribution to collected photons of Cherenkov light and scintillation
are the same for 120~GeV/$c$ electrons and for muons in the
5-20~GeV/$c$ region, within the 1.6\,\% precision of this measurement
(this last statement strictly holds only for muons at small angle with
respect to the crystal axis).

From the fit results, a muon critical energy of $160^{+5}_{-6}\,({\rm
 stat.}) \pm 8\,({\rm syst.})$~GeV is obtained, in agreement with the
computed value of 169.5~GeV for PbWO$_4$~\cite{ANP_Tables}. The
systematic uncertainty includes a contribution of 4.5~GeV from the
uncertainty of the containment corrections, dominated by the limited
knowledge of the correction for radiation losses, and a contribution
of 6~GeV due to the stability of the result against bias subtraction
and event selection. For the latter, sizable contributions come from
the variation of the acceptance of the muon angle with respect to the
crystal axis, and, to a lesser extent, from the requirement that $E/p$
be lower than 1. The contributions from these sources were estimated
by varying the muon acceptance angle between 0.3 and 0.7~radians, and
by removing the requirement that $E/p$ be lower than 1.\par

It might be argued that the use of pdf, calculated a priori in
anticipation of a particular outcome, would be inappropriate when the
fitted parameters differ from the expected values. In order to
investigate the possible importance of this, fits were performed with
alternative pdf sets, parametrized as functions of $\alpha$ and
$\beta$, with $\alpha$ controlling the pdf energy scale, and $\beta$
controlling the relative contributions of radiation and
collisions. The changes in the results were found to be negligible
compared to the quoted error.

\section{Conclusions}
\label{Conclusions}

The muon stopping power in $\textnormal{PbWO}_{4}$ has been measured
over the momentum range from 5~GeV/$c$ to 1000~GeV/$c$, yielding the
first experimental determination of the muon critical energy. The
value obtained: $160^{+5}_{-6}\,({\rm stat.}) \pm8.0\,({\rm
  syst.})$~GeV, agrees with the value calculated for this
material.\par

In the region corresponding to muon momenta less than 20~GeV/$c$, where
collision losses dominate, the average energy deposited in the
crystals is of order 300~MeV. Thus the agreement (to within 1-2\,\%)
between the measured stopping power and the calculated values confirms
that the energy calibration of the detector, previously determined
with 120~GeV/$c$ electrons, remains valid down to the sub-GeV
scale. Alternatively, the observed linearity in energy response may be
interpreted as demonstrating that the relative contributions of
Cherenkov and scintillation light to the measured signal are the same
for muons of $\sim$10~GeV/$c$ as for electrons of 120~GeV/$c$ (in the
specific case of muons traversing the crystals at small angles with
respect to the axis, and of electrons striking the crystals at normal
incidence). The confirmation of the detector energy calibration and
linearity of response relies on the validity of the calculated
stopping power in lead tungstate, in the region dominated by collision
losses.  On the other hand, the value extracted for the critical
energy does not depend on the energy calibration, although it would be
sensitive to a deviation from linearity of the response.

\section*{Acknowledgements}
We thank the technical and administrative staff at CERN and other CMS Institutes, and acknowledge support from: FMSR (Austria); FNRS and FWO (Belgium); CNPq, CAPES, FAPERJ, and FAPESP (Brazil); MES (Bulgaria); CERN; CAS, MoST, and NSFC (China); COLCIENCIAS (Colombia); MSES (Croatia); RPF (Cyprus); Academy of Sciences and NICPB (Estonia); Academy of Finland, ME, and HIP (Finland); CEA and CNRS/IN2P3 (France); BMBF, DFG, and HGF (Germany); GSRT (Greece); OTKA and NKTH (Hungary); DAE and DST (India); IPM (Iran); SFI (Ireland); INFN (Italy); NRF (Korea); LAS (Lithuania); CINVESTAV, CONACYT, SEP, and UASLP-FAI (Mexico); PAEC (Pakistan); SCSR (Poland); FCT (Portugal); JINR (Armenia, Belarus, Georgia, Ukraine, Uzbekistan); MST and MAE (Russia); MSTDS (Serbia); MICINN and CPAN (Spain); Swiss Funding Agencies (Switzerland); NSC (Taipei); TUBITAK and TAEK (Turkey); STFC (United Kingdom); DOE and NSF (USA). Individuals have received support from the Marie-Curie IEF program (European Union); the Leventis Foundation; the A. P. Sloan Foundation; and the Alexander von Humboldt Foundation.

\bibliography{auto_generated}   

\cleardoublepage\appendix\section{The CMS Collaboration \label{app:collab}}\begin{sloppypar}\hyphenpenalty=5000\widowpenalty=5000\clubpenalty=5000\textbf{Yerevan Physics Institute,  Yerevan,  Armenia}\\*[0pt]
S.~Chatrchyan, V.~Khachatryan, A.M.~Sirunyan
\vskip\cmsinstskip
\textbf{Institut f\"{u}r Hochenergiephysik der OeAW,  Wien,  Austria}\\*[0pt]
W.~Adam, B.~Arnold, H.~Bergauer, T.~Bergauer, M.~Dragicevic, M.~Eichberger, J.~Er\"{o}
% ...
\vskip\cmsinstskip
% ...

\end{sloppypar}
\end{document}